\documentclass[preprint,12pt,authoryear]{elsarticle}
\usepackage{graphicx,psfrag,epsf}
\usepackage{enumerate}
\usepackage{natbib}
\usepackage{url} 
\usepackage{amsmath,amssymb,amsfonts}
\usepackage{indentfirst}
\usepackage{algorithm, algorithmic}
\usepackage{booktabs} 
\usepackage{multirow}
\usepackage{tabularx}
\usepackage{amsthm}
\usepackage{color}
\usepackage{arydshln}
\usepackage{fancyhdr}
\usepackage{adjustbox}

\journal{ }

\begin{document}

\begin{frontmatter}



\title{Density Prediction of Income Distribution Based on Mixed Frequency Data}


\author[label1,label2]{Yinzhi Wang}
\author[label3]{Yingqiu Zhu}
\author[label4,label5]{Ben-Chang Shia}
\author[label3,label6]{Lei Qin\corref{correspondingauthor}}
\cortext[correspondingauthor]{Corresponding author at: School of Statistics, University of International Business and Economics, Beijing, China.}
\ead{qinlei@uibe.edu.cn}
\affiliation[label1]{organization={School of Finance, Shanghai University of International Business and Economics}
}
\affiliation[label2]{organization={Laboratory for Going Global Financial Services and Data Science, Shanghai University of International Business and Economics}
}
\affiliation[label3]{organization={School of Statistics, University of International Business and Economics}
}
\affiliation[label4]{organization={Graduate Institute of Business Administration, College of Management, Fu Jen Catholic University}
}
\affiliation[label5]{organization={Artificial Intelligence Development Center, Fu Jen Catholic University}
}
\affiliation[label6]{organization={Dong Fureng Institute of Economic and Social Development, Wuhan University}
}

\begin{abstract}
Modeling large dependent datasets in modern time series analysis is a crucial research 
area. One effective approach to handle such datasets is to transform the observations 
into density functions and apply statistical methods for further analysis. Income 
distribution forecasting, a common application scenario, benefits from predicting 
density functions as it accounts for uncertainty around point estimates, leading to 
more informed policy formulation. However, predictive modeling becomes challenging 
when dealing with mixed-frequency data. To address this challenge, this paper 
introduces a mixed data sampling regression model for probability density functions 
(PDF-MIDAS). To mitigate variance inflation caused by high-frequency prediction 
variables, we utilize exponential Almon polynomials with fewer parameters to 
regularize the coefficient structure. Additionally, we propose an iterative 
estimation method based on quadratic programming and the BFGS algorithm. Simulation 
analyses demonstrate that as the sample size for estimating density functions and 
observation length increase, the estimator approaches the true value. 
Real data analysis reveals that compared to single-sequence prediction models, 
PDF-MIDAS incorporating high-frequency exogenous variables offers a wider range of 
application scenarios with superior fitting and prediction performance. 
\end{abstract}


\end{frontmatter}


\section{Introduction}
With ongoing economic system reforms in various countries, there is growing attention 
on increasing residents' income and improving the distribution system. Over the past 
decade, notable progress has been made, such as the per capita disposable income of 
Chinese residents rising from 16500 to 35100. However, significant 
disparities persist in urban-rural development and income distribution in China, 
highlighting various imbalances that need to be addressed. To achieve effective 
economic development, it is essential to broaden the perspective from focusing solely 
on the "mean" to considering the entire income "distribution." This entails seeking 
ways to increase residents' income while simultaneously narrowing the wealth disparity 
between the rich and the poor. Accomplishing this goal necessitates enhancing the 
income distribution system and fostering the growth of the middle-income segment. 
Consequently, forecasting research on household income distribution holds significant 
practical significance as it provides valuable data-driven insights to support 
decision-making aimed at promoting shared prosperity.  

Household income serves as a prominent indicator in quantitative economic analysis. 
At the micro level, it reflects the practical 
purchasing power and living standards of residents. On a macro scale, household income 
distribution is a crucial measure of socioeconomic status and distribution fairness 
(Smeeding and Weinberg, 2001). Empirical analysis related to economics and people's 
livelihood often focuses on household income, examining its relationship with 
consumption and saving behavior (Zhou et al., 2009), the impact of family 
income on the physical fitness of adolescents (Ali et al., 2011; Murasko, 2013), 
the connection between agricultural development and household income (Noltze et al., 
2013), and the influence of household income structure on financial asset allocation 
(Zhang et al., 2015). Numerous studies have highlighted the significant disparities in 
household income distribution between urban and rural areas, different regions, and 
various social groups (Khan and Riskin, 2005; Zou and Wang , 2011; Xia et al., 2012; 
Khan et al., 2017). According to data from the National Bureau of Statistics of China, 
the per capita disposable income of household in the low-income group\footnote{The 
national household income is distributed among five groups, with the lowest 20\% of 
households categorized as the low-income group and the top 20\% of households 
classified as the high-income group.} was 8601 yuan 
in 2022, while that of the high-income group was 90116 yuan, indicating a difference 
of over 10 times, as depicted in Figure 1(a). Figure 1(b) displays the annual change 
rate of per capita disposable income for each income group from 2018 to 2022 compared 
to the previous year. Figure 1(b) reveals that the per capita income of 
households at different income levels exhibits diverse trends over time. For instance, 
low-income families experienced substantial income growth in 2018 and 2019 due to 
policy support. However, in 2021, the COVID-19 pandemic led to a sharp decline in 
income growth for low-income households. The evolution of household income 
distribution represents a complex and dynamic time series analysis challenge. 
For formulating policies related to income distribution, accurate predictions of household 
income distribution can provide essential data support. This necessitates the 
development of accurate and reliable quantitative models for household income 
distribution.
\begin{figure}[H]
    \begin{minipage}{0.5\linewidth}
    \centering
    \includegraphics[width=2.7in]{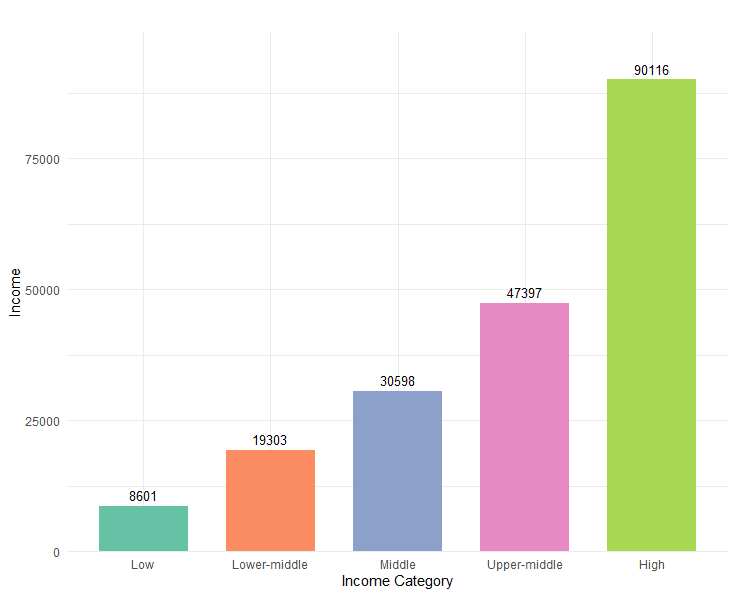}\\
    { (a)  }
    \end{minipage}
    \begin{minipage}{0.5\linewidth}
    
    \centering
    \includegraphics[width=2.7in]{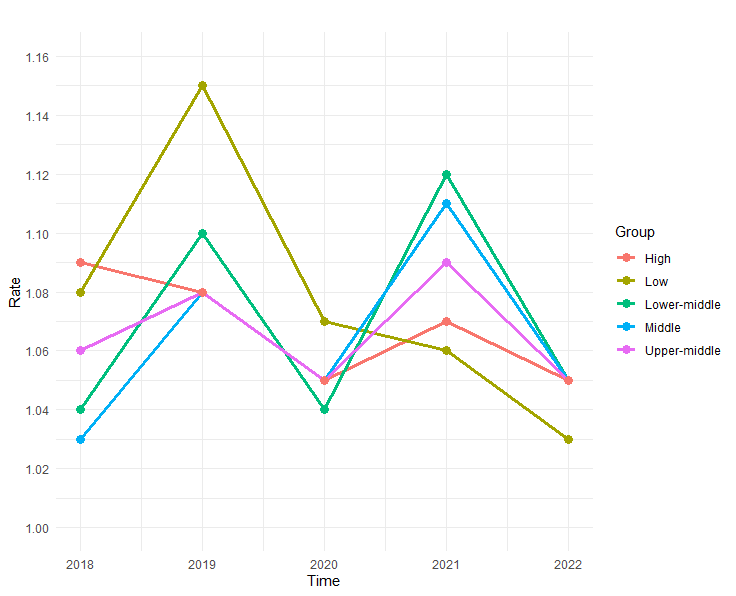}\\
    { (b)  }
    \end{minipage}
    \caption{(a): Per capita disposable income of households in five income quintiles of national residents in 2022. 
    (b): Change rate of per capita disposable income of households in each group relative to the previous year.}
\end{figure}

The primary challenge in predicting household income distribution is modeling time 
series using distribution data. Traditional time series analysis relies 
on point estimates and forecasts based on those estimates, such as autoregressive 
moving average models (Box et al., 2015), generalized autoregressive conditional 
heteroskedasticity models (Bollerslev, 1986), and quantile regression models (Koenker, 
2017). However, point estimates provide limited information and fail to fully capture 
the complete distribution of household income. Moreover, statistical inference based 
on point estimates often assumes time series stationarity and data normality, which 
may not hold true for actual economic distributions like household income, which 
exhibit characteristics such as time variability and non-normality. These factors 
limit the effectiveness of point estimation methods. This article focuses on 
predicting the probability density function (PDF) of household income distribution. 
Notable studies, such as the Hellinger distance autoregressive model (HDAR) by Tsay 
(2016) and the density function autoregressive model (FAR) by Chaudhuri et al. (2016), are 
discussed. Tsay (2016) suggests 
transforming multiple observations at the same time point into PDFs and subsequently 
conducting statistical modeling. The approach involves using a linear combination, 
where non-negative parameters that sum to $1$ are employed to weight and aggregate the 
PDFs from multiple lag periods. This transformation offers the advantage of enabling 
statistical inference using functional data methods (Ramsay and Silverman, 2005; Yao 
et al., 2005a, 2005b; Horváth and Kokoszka, 2012; Wang et al., 2016). 
Chaudhuri et al. (2016) employed the autoregressive operator to transform the lagged 
PDF for approximating the current period's PDF. This approach shares similarities 
with the methodologies proposed by Bosq (2000), Cardot et al. (1999), and Park and 
Qian (2007, 2012). The FAR model by Chaudhuri et al. (2016) offers significant 
potential for further exploration of PDFs. Subsequently, 
Chen et al. (2019) analyzed functional changes in liquidity supply and demand in the 
limit order book, while Cai et al. (2019) introduced FARVaR, a functional calculation 
method for daily value at risk (VaR). However, the FAR model has limitations as it 
solely relies on a single-sequence prediction method and does not take into account 
the impact of exogenous variables. Furthermore, FAR is suitable for modeling 
continuous time series, whereas household income distribution is typically represented 
by discontinuous time series. Hence, modeling discontinuous distributed data presents 
a challenge that must be addressed. This work aims to construct a 
comprehensive model by incorporating the relationship between household income and 
external variables. 

The observation frequencies of many indicators that affect household income are not 
uniform. Integrating multi-source and mixed-frequency observation information presents 
a challenge in predicting income distribution. Several factors, including taxes (Auten 
and Carroll, 1999), residents' employment status (Dynan et al., 2012), education level 
(Zhou Xuejiao and Liu Hefei, 2022), macroeconomic environment changes (Fallon and 
Lucas, 2002), and government fiscal expenditure changes (Tang Gaojie et al., 2023), 
impact household income. However, while household income is typically observed 
annually, other indicators as exogenous variables are usually collected on a monthly 
or quarterly basis. Existing literature on mixing sequences generally falls into three 
categories of processing methods. The first is frequency alignment methods, which 
involve reducing high-frequency observation sequences to low-frequency ones by either 
deleting or summing data points. This approach often results in the loss of 
valuable information. The second is the missing completion method, where low-frequency 
observation sequences are transformed into high-frequency ones through interpolation 
or splitting. However, determining the optimal interpolation method in advance is 
challenging, and measurement errors can easily affect the calculation results. The 
third is mixing sequence modeling methods, which involve combining high-frequency 
observation sequences and then performing regression predictions 
on low-frequency sequences. This model does not discard any 
information and has gained significant attention in empirical analyses. 
The mixed data sampling (MIDAS) model, introduced by Ghysels et al. (2004), is a 
renowned approach for modeling mixed-frequency sequences. This model constructs 
high-dimensional autoregressive models by utilizing low-frequency data as dependent 
variables and lagged high-frequency data as independent variables. The high 
dimensionality arises from the inclusion of high-frequency observations and their lag 
terms. To mitigate parameter inflation, the MIDAS model simplifies the high-dimensional 
influence coefficients using a weight function controlled by a limited number of 
parameters, such as exponential polynomial or Beta function. Within the MIDAS modeling 
framework, several extensions have been proposed. Ghysels et al. (2007) introduced the 
generalized MIDAS model, Engle et al. (2013) developed the GARCH-MIDAS model, and 
Guérin and Marcellino (2013) proposed the Markov-Switching MIDAS model. These models 
have shown success in predicting economic phenomena like the growth rate and provide 
valuable insights for research on income distribution prediction. 
However, the aforementioned models primarily focus on 
one-dimensional or low-dimensional time series indicators and do not explicitly 
consider the time series analysis of PDFs. 

This work proposes a mixed data sampling 
regression model for probability density functions (PDF-MIDAS) that 
addresses the prediction of household income distribution. The contributions of this 
work can be summarized as follows. First, a novel time series 
analysis method for PDFs is proposed. In the PDF-MIDAS model, both the dependent and 
independent variables are treated as PDFs. Second, multiple mixed sampling variables 
are incorporated into the model. Building upon the MIDAS model, this work proposes a 
simplified parameter structure and employs nonlinear optimization to estimate the 
parameters. This method is not only applicable for predicting income distribution 
but also provides valuable insights for analyzing distribution functions in other 
time series contexts involving mixing sequences. Third, the PDF-MIDAS model exhibits 
strong predictive capability for discontinuous time series data, assuming the 
availability of the independent variables. 

The remainder of this paper is organized as follows. The mixed data sampling 
regression model for probability density functions is introduced in Section 2. Section 
3 outlines the specific optimization process for estimating the model parameters. Simulation 
results are provided in Section 4, and a real-data example is presented in Section 5. 
Finally, Section 6 concludes this work. 
\section{Model}
\subsection{Univariate PDF-MIDAS model}
We consider two time series consisting of observed individuals represented as PDFs. 
The first is the dependent variable $f_{t}(x)$, $t=1,\ldots,T$, which represents the 
PDF of the annual observation. The second is the density $g_{t}^{(m)}(x)$ of the exogenous variable, 
where $m$ is the observation frequency. Specifically, the variable can be observed $m$ 
times from point $t-1$ to $t$, e.g., $m=12$ for monthly data. Referring to Tsay (2016), 
the $f_{t}(x)$ can be expressed as a combination of $g_{t}^{(m)}(x)$ and its 
$p$ lag terms, 
\begin{equation}
    f_{t}(x)=\sum_{i=1}^{p}c_{i}g_{t-h-i/m}^{(m)}(x)+e_{t}(x),
\end{equation}
where, $h$ represents the minimum interval period between the independent variable and 
the dependent variable. The lag time of high-frequency observation $g_{t}^{(m)}(x)$ is 
expressed as a fraction $i/m$, $i=1,\ldots,p$. The weight coefficient $c_{i}$ satisfies 
$c_{i}>0$ and $\sum_{i=1}^{p}c_{i}=1$, ensuring that the prediction result remains a PDF. $e_{t}(x)$ 
is the residual function, satisfying $\int e_{t}(x)dx=0$. Referring to Tsay (2016), 
equation (1) represents an extension of the HDAR model. The result of summarizing the 
PDFs using coefficients that sum to 1 remains a valid PDF. Furthermore, equation (1) 
simplifies the FAR model by transforming the 
autoregressive operator from a square matrix into a single parameter $c_{i}$. This 
simplification addresses the challenge of low prediction accuracy resulting from the 
poor estimation of the autoregressive operator. 

Due to the presence of high-frequency independent variables, the $p$ in equation (1) 
can potentially become large, leading to a high-dimensional regression model. For 
instance, certain monthly observation data may have over 20 observations within a 
two-year lag. The inclusion of numerous lag variables increases the complexity of the model. 
To address the issue of parameter expansion, this paper takes inspiration 
from the weight function employed in the MIDAS model (Ghysels et al., 2004). By 
simplifying the parameter 
structure, the paper proposes a univariate mixed data sampling regression model for  
probability density functions, referred to as PDF-MIDAS(1), 
\begin{equation}
    f_{t}(x)=B(L^{1/m},\Theta)g_{t-h}^{(m)}(x)+e_{t}(x),
\end{equation}
where $B(L^{1/m},\Theta)=\sum_{i=1}^{p}b(i,\Theta)L^{1/m}$ is the polynomial of the 
lag operator $L$, satisfying $L^{1/m}g_{t-h}^{(m)}(x)=g_{t-h-i/m}^{(m)}(x)$. 
The distinction between (2) and (1) lies in the coefficients of high-frequency 
variables, which exhibit multiple variations and are limited by the parameter 
structure $b(i,\Theta)$. A commonly employed weight structure is the exponential Almon 
polynomial, where $\Theta=(\theta_{1},\ldots,\theta_{q})$, 
\begin{equation}
    b(i,\Theta)=\frac{{\rm{exp}}(\theta_{1}i+\cdots+\theta_{q}i^{q})}{\sum_{j=1}^{p}{\rm{exp}}(\theta_{1}j+\cdots+\theta_{q}j^{q})}.
\end{equation}
where, $\sum_{i=1}^{p}b(i,\Theta)=1$. For example, when $q=1$ and $\theta_{1}=-1$, 
$b(i,\Theta)={\rm{exp}}(-i)/\sum_{j=1}^{p}{\rm{exp}}(-j)$ decreases as the lag order 
increases. The exponential Almon polynomial exhibits the property of gradually 
decreasing to zero as the lag term increases. This characteristic aligns with the 
typical decay pattern observed in time series analysis. In a similar vein, Ghysels et 
al. (2007) mentioned the Beta polynomial, which also captures the decay process of 
the weight coefficient using a small number of parameters. However, the Beta 
polynomial involves the use of the Beta function, making its estimation more 
challenging. As a result, the exponential Almon polynomial has emerged as the 
preferred choice for weight setting in practical applications. 
\subsection{Multivariate PDF-MIDAS model}
We consider multiple time series consisting of observed individuals represented as PDFs. 
The density of different exogenous variables is denoted as $g_{t,k}^{(m_{k})}(x)$, 
$k=1,\ldots,K$, where $m_{k}$ is the sampling frequency of the $k$th high-frequency 
independent variable. The multivariable model is a natural extension of the 
univariable model, expressing $f_{t}(x)$ as a linear combination of the 
$g_{t,k}^{(m_{k})}(x)$ and its $p_{k}$ lag terms, 
\begin{equation}
    f_{t}(x)=\sum_{k=1}^{K}a_{k}\left(\sum_{i=1}^{p_{k}}c_{i,k}g_{t-h-i/m_{k}}^{(m_{k})}(x)\right)+e_{t}(x),
\end{equation}
where, $c_{i,k}$ is the weight coefficient of each independent variable lag term, 
satisfying $c_{i,k}>0$ and $\sum_{i=1}^{p_{k}}c_{i,k}=1$. Since the linear combination 
$\sum_{i=1}^{p_{k}}c_{i,k}g_{t-h-i/m_{k}}^{(m_{k})}(x)$ is still a PDF, $a_{k}$ is 
used to summarize these combination structures, requiring $a_{k}>0$ and 
$\sum_{k=1}^{K}a_{k}=1$. Similarly, this paper proposes a multivariable mixed data 
sampling regression model for probability density function (PDF-MIDAS(K)) based on 
Almon weight polynomial,
\begin{equation}
    f_{t}(x)=\sum_{k=1}^{K}a_{k}B_{k}(L^{1/m_{k}},\Theta_{k})g_{t-h,k}^{(m_{k})}(x)+e_{t}(x),
\end{equation}
where, $B_{k}(L^{1/m_{k}},\Theta_{k})=\sum_{i=1}^{p_{k}}b(i,\Theta_{k})L^{1/m_{k}}$ is 
the lag operator polynomial of the $k$th high-frequency independent variable, and the 
influencing parameter is $\Theta_{k}=(\theta_{k,1},\ldots,\theta_{k,q_{k}})$. 
\section{Estimation and Property}
This section outlines the parameter estimation process of the PDF-MIDAS model, which 
includes the estimation of the PDF, setting of the objective 
function, parameter solution, and the asymptotic property of the estimator. 
\subsection{Parameter Estimation}
First, we employ the kernel density method to estimate the PDF of the variables. 
Let $\{x_1,\ldots,x_{M}\}$ denote a simple random sample of a random variable 
$X$. The kernel density estimate of its PDF can be represented as 
\begin{equation}
    {\hat{f}}(x)=\frac{1}{Ml}\sum_{i=1}^{M}K\left(\frac{x-x_{i}}{l}\right),
\end{equation}
where, $K(z)$ is the kernel function, satisfying symmetry, $\int K(z)dz=1$ and 
${\rm{lim}}_{z\rightarrow \infty}K(z)={\rm{lim}}_{z\rightarrow 0}K(z)=0$. In this work, 
we adopt the classic Gaussian kernel function $K(z)=(1/\sqrt{2\pi}){\rm{exp}}(-z^2/2)$. 
$l$ represents the window width of the kernel function. According to Tsay (2016), it 
is common to choose $l=0.9{\rm{min}}({\hat{\sigma}},{\rm{IQR}}/1.34)n^{-0.2}$, where 
${\hat{\sigma}}$ represents the standard deviation and ${\rm{IQR}}$ represents the 
quartile deviation of the sample. We estimate the density function using $N$ equidistant 
points $\{s_{1},\ldots,s_{N}\}$. In this paper, $N=30$, and the interval between 
consecutive points is denoted as $\Delta s$. Accordingly, we can represent the density 
function $f(x)$ using $N\times 1$ dimensional estimation results. 

Next, we introduce the objective function used for parameter estimation. On the shared 
interval $[\delta_{1}, \delta_{2}]$, there exist various types of distances $D(f,g)$ 
between two densities $f(x)$ and $g(x)$. Examples of such distances include the 
$L_{1}$ distance $\|f-g\|_{1}=\int |f(x)-g(x)|dx$, $L_{2}$ distance $\|f-g\|_{2}=
\sqrt{\int |f(x)-g(x)|^{2}dx}$, $L_{\infty}$ distance $\|f-g\|_{\infty}={\rm{sup}}|f(x)-
g(x)|dx$, and Hellinger distance $h(f,g)=\int(\sqrt{f(x)}-\sqrt{g(x)})^{2}dx$. 
Once the distance metric is chosen, for the PDF-MIDAS(1) model, the objective function 
can be expressed as 
\begin{equation}
Q(\Theta)=\sum_{t=1}^{T}D\left(f_{t}(x),B(L^{1/m},\Theta)g_{t-h}^{(m)}(x)\right).
\end{equation}
Similarly, for the PDF-MIDAS(K) model, the objective function can be expressed as
\begin{equation}
    Q(\Theta_{1},\ldots,\Theta_{K},a_{1},\ldots,a_{K})=\sum_{t=1}^{T}D\left(f_{t}(x),\sum_{k=1}^{K}a_{k}B_{k}(L^{1/m_{k}},\Theta_{k})g_{t-h,k}^{(m_{k})}(x)\right).
\end{equation}

Finally, quadratic optimization and nonlinear optimization techniques are utilized to 
iteratively solve the objective function. Specifically, we focus on solving the 
PDF-MIDAS(K) model using the $L_{2}$ distance. The other situations are similar. In 
this case, the objective function can be expressed as 
\begin{equation}
    Q(\Phi)=\sum_{t=1}^{T}\sum_{i=1}^{N}
    \left(f_{t}(s_{i})-\sum_{k=1}^{K}a_{k}B_{k}(L^{1/m_{k}},\Theta_{k})g_{t-h,k}^{
    (m_{k})}(s_{i})\right)^2\Delta s,
\end{equation}
where, $\Phi=(\Theta_{1},\ldots,\Theta_{K},a_{1},\ldots,a_{K})$. In equation (9), 
given $(\Theta_{1},\ldots,\Theta_{K})$, the solution for 
$(a_{1},\ldots,a_{K})$ can be reformulated as a classic quadratic optimization problem. 
Similarly, once given $(a_{1},\ldots,a_{K})$ the solution for $(\Theta_{1},\ldots,\Theta_{K})$ 
can be obtained using the BFGS algorithm. The BFGS algorithm is a popular method for 
solving optimization problems, particularly in the context of the MIDAS model. 
Then the above two steps are iterated until convergence, and the estimation result 
$({\hat{\Theta}}_{1},\ldots,{\hat{\Theta}}_{K},{\hat{a}}_{1},\ldots,{\hat{a}}_{K})$ 
can be obtained. 
\subsection{Asymptotic property} 
This subsection presents the asymptotic property of nonlinear least squares (NLS) estimation. 
For $\Phi=(\Theta_{1},\ldots,\Theta_{K},a_{1},\ldots,a_{K})$, we have 
\begin{equation}
    {\hat{\Phi}}={\rm{argmin}}_{\Phi}\sum_{t=1}^{T}\sum_{i=1}^{N}q_{ti}(\Phi),
\end{equation}
where, $q_{ti}(\Phi)=(f_{t}(s_{i})-\sum_{k=1}^{K}a_{k}B_{k}(L^{1/m_{k}},\Theta_{k})g_{t-h,k}^{
(m_{k})}(s_{i}))^2\Delta s$. Let $g_{t,i}=(g_{t-h-1/{m_{1}},1}^{(m_{1})}(s_{i}),
\ldots,g_{t-h-p_{K}/m_{K},K}^{(m_{k})}(s_{i}))$. To obtain the theoretical properties 
of parameter ${\hat{\Phi}}$, we make the 
following assumptions. 

{\textbf{Assumption 1:}} Each parameter in $\Phi$ is bounded, and for model (5) we have 
$E(f_{t}(x)|\Theta_{k},a_{k},k=1,\ldots,K)=\sum_{k=1}^{K}a_{k}B_{k}(L^{1/m_{k}},\Theta_{k})g_{t-h,k}^{(m_{k})}(x)$.

{\textbf{Assumption 2:}} For any $k=1,\ldots,K$, $\{(e_{t}(s_{i}),g_{t-h,k}^{(m_{k})}(s_{i}))\}$ is strong mixing.

{\textbf{Assumption 3:}} For any small $\varepsilon>0$, there exists an $r=8+\varepsilon$, 
such that the following equation holds. 
\[
   (1) \ E\|g_{t,i}\|_{s}^{r}\le C_{1}, \ (2) \ E\|e_{t}(s_{i})\|_{s}^{r}\le C_{2}, \ 
   (3) \ E\|e_{t}(s_{i})\|_{s}^{r}\|e_{t-h}(s_{i})\|_{s}^{r}\le C_{3}, 
\]
\[
  (4) \ E\|g_{t,i}\|_{s}^{r}\|g_{t-h,i}\|_{s}^{r}\le C_{4}, \ (5) \ E\|g_{t,i}\|_{s}^{r}\|e_{t-h}(s_{i})\|_{s}^{r}\le C_{5},  
\]
where, $\|\cdot\|_{s}$ is the vector norm defined by Mira and Escribano (1995). The 
$e_{t}(s_{i})$, $t=1,\ldots,T$, $i=1,\ldots,N$, are independent of each other. 

Assumption 1, Assumption 2 and Assumption 3 correspond to Assumption MD, Assumption MX and Assumption LF in Mira and Escribano (1995), respectively. 
Assumption 1 restricts the parameter space and assumes that the model is correctly 
specified. Assumption 2 allows for long-term dependence in the model and does not 
impose strict constraints on model heterogeneity. Assumption 3 restricts the moment 
conditions of $g_{t,i}$ and $e_{t}(s_{i})$, and assumes that the error $e_{t}(s_{i})$ 
at each grid point and time are independent of each other. Referring to Mira and 
Escribano (1995), we can get the following theorem. 
\newtheorem{theorem}{Theorem}
\begin{theorem}
    Based on Assumption 1-3, the NLS estimator ${\hat{\Phi}}$ has asymptotic normality. 
    \[
       (B)^{-0.5}A(\Phi)\sqrt{NT}\left({\hat{\Phi}}-\Phi\right)\sim N(0,I_{\sum_{k=1}^{K}q_{k}+K}), 
    \] 
    where, $B={\rm{Var}}((NT)^{-0.5}\sum_{t=1}^{T}\sum_{i=1}^{N}M_{it})$, $M_{it}=
    \triangledown_{\Phi}q_{ti}(\Phi)$. $A(\Phi)=\triangledown_{\Phi}^{2}{\bar{Q}}(\Phi)$, 
    where ${\bar{Q}}(\Phi)=E(Q(\Phi))$. And $I_{k}$ represents the identity matrix with dimension $k\times k$.      
\end{theorem}
\noindent
Theorem 1 demonstrates the asymptotic normality of the NLS estimation ${\hat{\Phi}}$. 
According to the theorem, as the observation length $T$ tends to infinity while the number of 
grid points $N$ remains fixed, the NLS estimator ${\hat{\Phi}}$ converges to the true 
value ${\Phi}$. Theorem 1 corresponds to Theorem 4.1 in Mira and Escribano (1995). 
The proof of Theorem 1 is provided in the Appendix. 
\section{Simulation}
\subsection{Simulation of univariate model}
The data generation process of the univariate model is shown in equation (2), where 
the observation length $T=\{100,200,500,1000\}$. The Almon polynomial $b(i,\Theta)$ in 
(3) considers two cases, the $q=1$ single-parameter form $\Theta=\theta_{1}=-0.05$ and the 
$q=2$ two-parameter form $\Theta=(\theta_{1},\theta_{2})=(0.2,-0.03)$. Sampling frequency 
$m=3$, minimum interval order $h=1/3$, and lag oeder $p$ takes the value from $\{3,12\}$. 
Let $g_{t-i/m}^{(m)}(x)$ follow the normal distribution $N(0.01t+i/m,1)$. According to (2), 
we obtain the expression of $f_{t}(x)$. Then, $M=\{100,500,1000\}$ points are 
sampled for both the dependent and independent variables at each time point to 
generate a set of simulation data. The Accept/Reject method in Casella and Berger 
(2002) can be employed to extract samples from the PDF of the dependent variable. 
Parameter estimation is performed according to the estimation process in Section 3.1. 
A total of 100 simulations were conducted. The evaluation criteria for assessing the 
estimated effect are represented by bias, standard deviation (SD), and root mean 
square error (RMSE). For example, for $\theta_{1}$, 
\[
   {\rm{Bias}}= \sum_{i=1}^{100}({\hat{\theta}}_{1}^{(i)}-{\theta}_{1})/100, \ {\rm{SD}}=\sqrt{\sum_{i=1}^{100}({\hat{\theta}}_{1}^{(i)}-{\bar{\theta}}_{1})^{2}/100},
\]
\[
    {\rm{RMSE}}=\sqrt{\sum_{i=1}^{100}({\hat{\theta}}_{1}^{(i)}-{\theta}_{1})^{2}/100},
\]
where, ${\hat{\theta}}_{1}^{(i)}$ represents the estimation result produced by the 
$i$th simulation sample and ${\bar{\theta}}_{1}=\sum_{i=1}^{100}{\hat{\theta}}_{1}^{(i)}/100$.

Tables 1 to 3 present the parameter estimation results for 100 sets of simulated data 
when the $M$ is $100$, $500$, or $1000$, respectively. Based on these results, 
the following conclusions can be drawn. First, the number of $M$ at each time point 
and the observation length $T$ jointly determine the estimation effectiveness of the 
PDF. Increasing both $M$ and $T$ simultaneously can lead to a lower RMSE of the 
estimator. This decrease in RMSE suggests that the estimator can be close to the true 
value. Second, when $M$ is fixed, increasing the $T$ generally leads to a decrease in 
both the RMSE and the SD of the estimator, regardless 
of whether the weight function $b(i,\Theta)$ has a single parameter or two parameters. 
However, when $M$ is small, the bias may still be large even with increasing $T$. For 
example, when $M=100$ and $p=12$, the bias of the estimator $\theta_{1}$ remains 
relatively stable at a certain level without a significant decrease. Third, 
when the $T$ is fixed, increasing the $M$ leads to a clear downward trend in the 
bias, SD, and RMSE of the estimator, irrespective of whether the weight function has a 
single parameter or two parameters.
For instance, consider the scenario where $q=1$ and $p=12$. As the $M$ increases, the 
estimated SD and RMSE of the parameter $\theta_{1}$ demonstrate a substantial decrease. 
Finally, when both $T$ and $M$ are fixed, increasing the lag order $p$ results in a 
notable improvement in the estimation effect. For example, when $M=1000$, $T=100$, and 
$q=2$, as the lag order $p$ increases, the 
estimators for $\theta_{1}$ and $\theta_{2}$ exhibit a substantial decrease in bias, 
SD, and RMSE. 
\begin{table}
    \centering
    \caption{Parameter estimation results of PDF-MIDAS(1) model when $M=100$}
    \begin{adjustbox}{width=\textwidth}
    \begin{tabular}{cccccccc}
    \toprule
        &  & $p=3$ & $p=12$ & $p=3$ &  
        & $p=12$ &  \\
    \midrule
        &  & $\theta_{1}=-0.05$ & $\theta_{1}=-0.05$ & $\theta_{1}=0.2$ 
        & $\theta_{2}=-0.03$ & $\theta_{1}=0.2$ & $\theta_{2}=-0.03$ \\
    \midrule
        $T=100$ & Bias &0.0043 & 0.0024 & -0.2630 & 0.0602 & -0.0784 & 0.0076\\
        & SD &0.0525 & 0.0042 & 0.7359 & 0.1795 & 0.0371 & 0.0032\\
        & RMSE &0.0527 & 0.0048 & 0.7815 & 0.1893 & 0.0867 & 0.0083\\
        $T=200$ & Bias &0.0138 & 0.0027 & -0.2326 & 0.0533 & -0.0767 & 0.0074\\
        & SD &0.0351 & 0.0029 & 0.5578 & 0.1359 & 0.0253 & 0.0022\\
        & RMSE &0.0377 & 0.0039 & 0.6043 & 0.1460 & 0.0807 & 0.0078\\
        $T=500$ & Bias &0.0090 & 0.0026 & -0.2193 & 0.0488 & -0.0762 & 0.0074\\
        & SD &0.0266 & 0.0017 & 0.3502 & 0.0870 & 0.0166 & 0.0015\\
        & RMSE &0.0281 & 0.0031 & 0.4132 & 0.0997 & 0.0780 & 0.0075\\
        $T=1000$ & Bias &0.0088 & 0.0025 & -0.2384 & 0.0546 & -0.0754 & 0.0073\\
        & SD &0.0189 & 0.0013 & 0.2591 & 0.0641 & 0.0122 & 0.0011\\
        & RMSE &0.0209 & 0.0028 & 0.3521 & 0.0842 & 0.0764 & 0.0074\\
    \bottomrule
    \end{tabular}
    \end{adjustbox}
\end{table}
\begin{table}
    \centering
    \caption{Parameter estimation results of PDF-MIDAS(1) model when $M=500$}
    \begin{adjustbox}{width=\textwidth}
    \begin{tabular}{cccccccc}
    \toprule
        &  & $p=3$ & $p=12$ & $p=3$ &  
        & $p=12$ &  \\
    \midrule
        &  & $\theta_{1}=-0.05$ & $\theta_{1}=-0.05$ & $\theta_{1}=0.2$ 
        & $\theta_{2}=-0.03$ & $\theta_{1}=0.2$ & $\theta_{2}=-0.03$ \\
    \midrule
        $T=100$ & Bias &0.0033 & 0.0010 & -0.3382 & 0.0825 & -0.0431 & 0.0041\\
        & SD &0.0315 & 0.0017 & 0.7118 & 0.1751 & 0.0188 & 0.0016\\
        & RMSE &0.0317 & 0.0019 & 0.7880 & 0.1936 & 0.0470 & 0.0045\\
        $T=200$ & Bias &0.0001 & 0.0013 & -0.3163 & 0.0777 & -0.0415 & 0.0040\\
        & SD &0.0206 & 0.0014 & 0.4412 & 0.1079 & 0.0129 & 0.0011\\
        & RMSE &0.0206 & 0.0019 & 0.5429 & 0.1330 & 0.0434 & 0.0042\\
        $T=500$ & Bias &0.0062 & 0.0013 & -0.2932 & 0.0717 & -0.0408 & 0.0040\\
        & SD &0.0139 & 0.0009 & 0.3099 & 0.0764 & 0.0085 & 0.0007\\
        & RMSE &0.0152 & 0.0016 & 0.4266 & 0.1048 & 0.0416 & 0.0040\\
        $T=1000$ & Bias &0.0040 & 0.0013 & -0.2902 & 0.0708 & -0.0397 & 0.0039\\
        & SD &0.0101 & 0.0007 & 0.2235 & 0.0552 & 0.0060 & 0.0005\\
        & RMSE &0.0109 & 0.0015 & 0.3663 & 0.0898 & 0.0402 & 0.0039\\
    \bottomrule
    \end{tabular}
    \end{adjustbox}
\end{table}
\begin{table}
    \centering
    \caption{Parameter estimation results of PDF-MIDAS(1) model when $M=1000$}
    \begin{adjustbox}{width=\textwidth}
    \begin{tabular}{cccccccc}
    \toprule
        &  & $p=3$ & $p=12$ & $p=3$ &  
        & $p=12$ &  \\
    \midrule
        &  & $\theta_{1}=-0.05$ & $\theta_{1}=-0.05$ & $\theta_{1}=0.2$ 
        & $\theta_{2}=-0.03$ & $\theta_{1}=0.2$ & $\theta_{2}=-0.03$ \\
    \midrule
        $T=100$ & Bias &-0.0005 & 0.0011 & -0.2969 & 0.0736 & -0.0315 & 0.0031\\
        & SD &0.0211 & 0.0013 & 0.6266 & 0.1544 & 0.0128 & 0.0011\\
        & RMSE &0.0211 & 0.0017 & 0.6934 & 0.1711 & 0.0340 & 0.0033\\
        $T=200$ & Bias &0.0047 & 0.0011 & -0.2583 & 0.0634 & -0.0310 & 0.0030\\
        & SD &0.0153 & 0.0010 & 0.4448 & 0.1103 & 0.0104 & 0.0009\\
        & RMSE &0.0161 & 0.0014 & 0.5143 & 0.1272 & 0.0327 & 0.0032\\
        $T=500$ & Bias &0.0024 & 0.0010 & -0.2413 & 0.0592 & -0.0312 & 0.0030\\
        & SD &0.0100 & 0.0006 & 0.2889 & 0.0716 & 0.0072 & 0.0007\\
        & RMSE &0.0103 & 0.0012 & 0.3764 & 0.0929 & 0.0320 & 0.0031\\
        $T=1000$ & Bias &0.0035 & 0.0010 & -0.2793 & 0.0685 & -0.0305 & 0.0030\\
        & SD &0.0063 & 0.0004 & 0.1932 & 0.0479 & 0.0053 & 0.0004\\
        & RMSE &0.0072 & 0.0011 & 0.3396 & 0.0836 & 0.0309 & 0.0030\\
    \bottomrule
    \end{tabular}
    \end{adjustbox}
\end{table}
\subsection{Simulation of multivariate model}
The data generation process of the multivariate model is shown in equation (5), where 
The number of high-frequency independent variables $K=2$. Let the number of parameters 
of the Almon weight function of the first and second high-frequency independent 
variables be $q_{1}=1$ and $q_{2}=2$ respectively, $\Theta_{1}=\theta_{1,1}=-0.05$ and 
$\Theta_{2}=(\theta_{2,1},\theta_{2,2})=(0.2,-0.03)$. The sampling frequency of two high-frequency variables is   
$m_{1}=m_{2}=3$, minimum interval order $h=1/3$, and lag order $p_{1}=p_{2}$ takes the 
value from $\{3,12\}$. Combination coefficient of independent variables $(a_1,a_2)=(0.4,0.6)$. 
Let $g_{t-i/m_{1},1}^{(m_1)}(x)$ follow the normal distribution $N(0.01t+i/m_{1},1)$ and 
$g_{t-i/m_{2},2}^{(m_2)}(x)$ follow the normal distribution $N(0.012t+i/m_{2},2)$. 
According to (5), 
we obtain the expression of $f_{t}(x)$. Then, $M=\{100,500,1000\}$ points are 
sampled for both the dependent and independent variables at each time point to 
generate a set of simulation data. The observation length $T$ takes the value from 
${100,200,500,1000}$. The Accept/Reject method is also employed to extract samples 
from the PDF of the dependent variable. 
Parameter estimation is performed according to the estimation process in Section 3.1. 
A total of 100 simulations were conducted. Similarly, we provide the bias, SD, and 
RMSE of the estimator based on 100 sets of simulated data. 
\begin{table}
    \centering
    \caption{Parameter estimation results of multivariate model when $M=100$}
    \begin{adjustbox}{width=\textwidth}
    \begin{tabular}{cccccccccc}
    \toprule
        &  & $p_{1}=p_{2}=3$ &  &  &  
        & $p_{1}=p_{2}=12$ & & &  \\
    \midrule
        &  & $a_{1}=0.4$ & $\theta_{1,1}=-0.05$ & $\theta_{2,1}=0.2$ 
        & $\theta_{2,2}=-0.03$ & $a_{1}=0.4$ & $\theta_{1}=-0.05$ & $\theta_{2,1}=0.2$ 
        & $\theta_{2,2}=-0.03$ \\
    \midrule
        $T=100$ & Bias &0.1154 & 0.3032 & 1.4808 & -0.2532 & 0.0408 & -0.0548 &0.1593 &-0.0067\\
        & SD &0.0412 & 0.2037 & 3.4045 & 0.7824 & 0.0750 & 0.0522 &0.2885 &0.0262\\
        & RMSE &0.1225 & 0.3653 & 3.7126 & 0.8224 & 0.0853 & 0.0757 & 0.3295 &0.0270\\
        $T=200$ & Bias &0.0486 & 0.1309 & 0.1802 & -0.0116 & -0.0199 & -0.0362 &-0.0015 &0.0050\\
        & SD &0.0266 & 0.1582 & 1.6170 & 0.3855 & 0.0450 & 0.0323 &0.1492 &0.0105\\
        & RMSE &0.0554 & 0.2053 & 1.6270 & 0.3857 & 0.0492 & 0.0485 &0.1492 &0.0116\\
        $T=500$ & Bias &0.0027 & 0.0793 & -0.1290 & 0.0386 & 0.0026 &-0.0143 &-0.0904 &0.0105\\
        & SD &0.0145 & 0.0905 & 0.7765 & 0.1875 & 0.0191 & 0.0121 &-0.0742 &0.0056\\
        & RMSE &0.0147 & 0.1203 & 0.7872 & 0.1914 & 0.0193 & 0.0187 &0.1170 &0.0118\\
        $T=1000$ & Bias &-0.0145 & -0.0350 & -0.0224 & 0.0111 & -0.0031 & -0.0113 &-0.0757 &0.0089\\
        & SD &0.0060 & 0.0488 & 0.5338 & 0.1331 & 0.0132 & 0.0079 &0.0548 & 0.0042\\
        & RMSE &0.0157 & 0.0600 & 0.5342 & 0.1335 & 0.0136 & 0.0138 &0.0935 &0.0098\\
    \bottomrule
    \end{tabular}
    \end{adjustbox}
\end{table}
\begin{table}
    \centering
    \caption{Parameter estimation results of multivariate model when $M=500$}
    \begin{adjustbox}{width=\textwidth}
    \begin{tabular}{cccccccccc}
    \toprule
        &  & $p_{1}=p_{2}=3$ &  &  &  
        & $p_{1}=p_{2}=12$ & & &  \\
    \midrule
        &  & $a_{1}=0.4$ & $\theta_{1,1}=-0.05$ & $\theta_{2,1}=0.2$ 
        & $\theta_{2,2}=-0.03$ & $a_{1}=0.4$ & $\theta_{1}=-0.05$ & $\theta_{2,1}=0.2$ 
        & $\theta_{2,2}=-0.03$ \\
    \midrule
        $T=100$ & Bias &0.1490 & 0.1964 & 2.9689 & -0.4899 & 0.0303 & -0.0774 &0.3041 &-0.0044\\
        & SD &0.0185 & 0.1407 & 3.6634 & 0.8166 & 0.0789 & 0.0304 &0.1398 &0.0856\\
        & RMSE &0.1502 & 0.2416 & 4.7154 & 0.9523 & 0.0845 & 0.0831 & 0.3347 &0.0857\\
        $T=200$ & Bias &0.0790 & 0.0102 & 1.6364 & -0.2683 & -0.0159 & -0.0442 &0.1173 &-0.0045\\
        & SD &0.0168 & 0.0915 & 2.2562 & 0.5219 & 0.0274 & 0.0163 &0.0660 &0.0046\\
        & RMSE &0.0808 & 0.0920 & 2.7872 & 0.5869 & 0.0317 & 0.0471 &0.1346 &0.0064\\
        $T=500$ & Bias &0.0153 & 0.0111 & 0.3703 & -0.0564 & 0.0026 &-0.0164 &0.0039 &0.0021\\
        & SD &0.0098 & 0.0529 & 0.8602 & 0.2045 & 0.0140 & 0.0084 &0.0508 &0.0038\\
        & RMSE &0.0182 & 0.0540 & 0.9365 & 0.2122 & 0.0142 & 0.0184 &0.0510 &0.0043\\
        $T=1000$ & Bias &-0.0038 & -0.0281 & 0.2819 & -0.0521 & -0.0018 & -0.0097 &-0.0062 &0.0021\\
        & SD &0.0043 & 0.0266 & 0.4730 & 0.1142 & 0.0112 & 0.0039 &0.0377 & 0.0028\\
        & RMSE &0.0057 & 0.0387 & 0.5506 & 0.1255 & 0.0114 & 0.0104 &0.0382 &0.0035\\
    \bottomrule
    \end{tabular}
    \end{adjustbox}
\end{table}
\begin{table}
    \centering
    \caption{Parameter estimation results of multivariate model when $M=1000$}
    \begin{adjustbox}{width=\textwidth}
    \begin{tabular}{cccccccccc}
    \toprule
        &  & $p_{1}=p_{2}=3$ &  &  &  
        & $p_{1}=p_{2}=12$ & & &  \\
    \midrule
        &  & $a_{1}=0.4$ & $\theta_{1,1}=-0.05$ & $\theta_{2,1}=0.2$ 
        & $\theta_{2,2}=-0.03$ & $a_{1}=0.4$ & $\theta_{1}=-0.05$ & $\theta_{2,1}=0.2$ 
        & $\theta_{2,2}=-0.03$ \\
    \midrule
        $T=100$ & Bias &0.1577 & 0.1397 & 3.8041 & -0.5947 & 0.0111 & -0.0876 &0.3782 &-0.0214\\
        & SD &0.0187 & 0.0916 & 3.7327 & 0.8501 & 0.0292 & 0.0209 &0.0911 &0.0062\\
        & RMSE &0.1588 & 0.1671 & 5.3295 & 1.0375 & 0.0312 & 0.0900 & 0.3890 &0.0222\\
        $T=200$ & Bias &0.0881 & -0.0734 & 2.7158 & -0.4563 & 0.0163 & -0.0478 &0.1582 &-0.0077\\
        & SD &0.0123 & 0.0732 & 2.4356 & 0.5328 & 0.0232 & 0.0126 &0.0538 &0.0037\\
        & RMSE &0.0890 & 0.1037 & 3.6480 & 0.7015 & 0.0284 & 0.0495 &0.1671 &0.0085\\
        $T=500$ & Bias &0.0226 & -0.0127 & 0.8137 & -0.1441 & 0.0019 &-0.0178 &0.0399 &-0.0012\\
        & SD &0.0097 & 0.0418 & 1.0048 & 0.2402 & 0.0122 & 0.0070 &0.0429 &0.0032\\
        & RMSE &0.0246 & 0.0436 & 1.2929 & 0.2801 & 0.0123 & 0.0192 &0.0586 &0.0034\\
        $T=1000$ & Bias &0.0005 & -0.0172 & 0.5563 & -0.1135 & -0.0003 & -0.0090 &0.0152 &0.0001\\
        & SD &0.0033 & 0.0263 & 0.5142 & 0.1243 & 0.0101 & 0.0027 &0.0360 & 0.0027\\
        & RMSE &0.0034 & 0.0314 & 0.7576 & 0.1683 & 0.0101 & 0.0094 &0.0391 &0.0027\\
    \bottomrule
    \end{tabular}
    \end{adjustbox}
\end{table}

Tables 4 to 6 present the parameter estimation results for 100 sets of simulated data 
when the $M$ is $100$, $500$, or $1000$, respectively. Similar to the univariate 
results, we can draw the following conclusions. First, Simultaneously increasing the 
number of samples $M$ and the observation length $T$ will result in a lower RMSE of 
the estimator, bringing it closer to the true value. For example, when $p_{1}=p_{2}=12$, 
the RMSE of the $a_{1}$ estimator decreases as $T$ and $M$ increase simultaneously. 
Second, when $M$ is fixed, 
increasing the $T$ generally leads to a reduction in 
both the bias, the SD and the RMSE of the estimator.  
For example, when $M=1000$ and $p_{1}=p_{2}=12$, the bias, SD, and RMSE of the 
estimator $\theta_{1}$ all decrease significantly as $T$ increases. Third, 
when the $T$ is relatively large, increasing the $M$ leads to a clear downward trend in the 
RMSE of the estimator.
For instance, consider the scenario where $T=1000$ and $p_{1}=p_{2}=12$. As the $M$ increases, the 
estimated RMSE of the parameter $a_{1}$ demonstrate a substantial decrease. 
Finally, when both $T$ and $M$ are fixed, increasing the lag order $p$ results in a 
notable improvement in the estimation effect. For example, when $M=1000$ and $T=1000$, 
as the lag order $p$ increases, the 
estimators for $\theta_{2,1}$ and $\theta_{2,2}$ exhibit a substantial decrease in bias, 
SD, and RMSE.
\section{Prediction of income distribution probability density function}
\subsection{Data description}
This section aims to validate the rationality of the method by predicting the distribution 
function of Chinese household income. The income data is 
sourced from the 2010-2020 China Family Panel Studies (CFPS), which is published by 
the China Social Sciences Survey Center at Peking University. The CFPS database 
conducts comprehensive surveys on the income status of Chinese households, with an 
average dataset size of around 13,000 observations per year. Thus, it serves as a 
reliable and effective data source for describing the income distribution among 
residents in China. For each annual time point, we employ the kernel density method to 
estimate the PDF of household income. Additionally, according to 
Zhou and Liu (2022) and Tang et al. (2023), education level, family size, household 
income structure, and government fiscal 
expenditure are important factors affecting residents' income. Regarding education 
level, this article utilizes the highest academic level of family members from the 
CFPS database as a representation. The highest academic level is categorized into 
eight levels according to a hierarchical standard: 1 representing illiterate/semi-literate, 
2 representing primary school, 3 representing junior high school, 4 representing high 
school/technical secondary school/technical school/vocational high school, 5 
representing junior college, 6 representing undergraduate, 7 representing master, and 
8 representing doctorate. Family size refers to the total number of individuals within a 
household. On the other hand, the family income structure signifies the proportion of 
wage income earned by a family in relation to its overall income. Regarding local 
government fiscal expenditures, we collected 
monthly fiscal expenditure data for 149 cities from the Wind database. (there are 
instances of missing data in certain months for some cities). 
Table 7 presents the descriptive statistical 
results for each variable. By utilizing the kernel density method, we can derive the 
annual PDFs for education level, family size, and income structure ratio, and the monthly PDF 
for local government fiscal expenditure. This article aims to utilize the PDFs of 
education level, family size, income structure ratio, and local government fiscal 
expenditure to predict the PDF of household income. 
\begin{table}
    \centering
    \caption{Descriptive statistical results of each variable. $25\%$ and $75\%$ 
    represent the lower quartile and upper quartile respectively.}
    \begin{adjustbox}{width=1.15\textwidth}
    \begin{tabular}{cccccccccc}
    \toprule
        Variable & Year & Frequency & Sample size & Min & $25\%$ 
        & Median & Mean & $75\%$ & Max  \\
    \midrule
       household income &2010 & annual & 14692 &0 & 3000 & 14000 & 22765 & 30000 & 800000\\ 
       &2011 & annual & 13041 & 0 & 5000 & 18000 & 24476 & 30000 & 5240000\\
       & 2012 &annual & 11842 & 1 & 11159 & 29381 & 43459 & 55197 & 3036046\\
       &2014 &annual & 13546 &0 &20000 & 36000 &50350 &60000 &4270560\\
       &2016 &annual &13842 &0 &20000 &40000 &59809 &70000 &8336000\\
       &2018 &annual &13835 &0 &20000 &47000 &65588 &80000 &6500000\\
       &2020 &annual &11097 &0 &30000 &50000 &80254 &100000 &5000000\\
       education level &2010 &annual &46519 &1 &2 &3 &2.64 & 3 &8\\
       &2011 &annual &46519 &1 &2 &3 &2.64 & 3 &8\\
       &2012 & annual &52804 &1 &1 &3&2.59 &3 &8\\
       &2014 & annual &57095 &1 &1 &2&2.68 &3 &8\\
       &2016 & annual &56868 &1 &1 &2&2.42 &3 &8\\
       &2018 & annual &58053 &1 &1 &2&2.47 &3 &8\\
       &2020 & annual &51073 &1 &1 &2&2.60 &4 &8\\
       family size &2010 & annual & 14797 &1.00 & 3.00 & 4.00 & 3.82 & 5.00 & 26.00\\ 
       &2011 & annual & 13127 &1.00 & 3.00 & 4.00 & 3.88 & 5.00 & 27.00\\
       &2012 &annual & 13315 &1.00 & 3.00 & 4.00 & 3.83 & 5.00 & 17.00\\
       &2014 &annual & 11946 &1.00 & 2.00 & 3.00 & 3.71 & 5.00 & 17.00\\
       &2016 &annual & 14019 &1.00 & 2.00 & 3.00 & 3.71 & 5.00 & 19.00\\
       &2018 &annual & 14218 &1.00 & 2.00 & 3.00 & 3.60 & 5.00 & 21.00\\
       &2020 &annual & 11620 &1.00 & 2.00 & 3.00 & 3.66 & 5.00 & 15.00\\
       Income structure ratio &2010 &annual &13919 &0.000 &0.289 &0.652 &0.580 & 0.935 &1.000\\
       &2011 &annual &12665 &0.000 &0.315 &0.695 &0.608 & 0.989 &1.000\\
       &2012 & annual &11842 &0.000 &0.000 &0.550 &0.485 & 0.917 &1.000\\
       &2014 & annual &12701 &0.000 &0.000 &0.682 &0.551 & 0.949 &1.000\\
       &2016 & annual &13982 &0.000 &0.000 &0.593 &0.511 & 0.906 &1.000\\
       &2018 & annual &14215 &0.000 &0.125 &0.667 &0.566 & 0.940 &1.000\\
       &2020 & annual&11614 &0.000 &0.082 &0.677 &0.567 & 0.952 &1.000\\
       fiscal expenditures & 2010 &monthly &1068 &1.25 &7.09 &11.12 &18.68 &21.67 &285.12\\
       & 2011 &monthly &1104 &1.06 &8.95 &14.74 &23.66 &27.59 &338.96\\
       & 2012 &monthly &1128 &0.00 &11.36 &18.15 &27.21 &32.20 &247.23\\
       & 2014 &monthly &1380 &0.42 &11.21 &20.00 &30.31 &38.26 &400.82\\
       & 2016 &monthly &1512 &0.54 &15.25 &26.79 &38.42 &46.73 &661.55\\
       & 2018 &monthly &1632 &0.50 &16.28 &28.41 &42.62 &51.02 &557.97\\
       & 2020 &monthly &1611 &0.03 &19.64 &34.63 &50.01 &59.31 &609.96\\
    \bottomrule
    \end{tabular}
    \end{adjustbox}
\end{table} 

Figure 2 displays the fitting diagram of the PDF for Chinese household income. To 
investigate the temporal changes in income distribution, this study constructs the fitted income 
distribution plot, ensuring that it includes at least the $5\%-95\%$ quantile of 
income for each year. Extreme values are not considered in this analysis. 
The results depicted in Figure 2 reveal that from 2010 to 2020, the income of Chinese 
households exhibited a consistent upward trend, leading to a rightward shift in the 
income distribution. Furthermore, the density of the peak in the middle-income group 
has decreased, while the area of the right tail has expanded. This indicates a 
progressive flattening of the income distribution over the years. These findings 
imply substantial improvements in the income situation of Chinese residents, 
particularly when focusing on the non-high-income groups.
Figure 2 clearly demonstrates a noticeable right-skew trend in the income distribution 
over the seven-year period. This signifies that the median household income is lower 
than the average income. Furthermore, due to the discontinuous nature of the sampled 
time series for household income distribution, many autoregressive time series models 
cannot be effectively applied. For instance, the FAR model proposed by Chaudhuri et 
al. (2016) can only handle continuous time series, limiting its applicability. In 
contrast, the PDF-MIDAS model proposed in this article can effectively handle 
discontinuous time series. The modeling process of PDF-MIDAS is elaborated below.
\begin{figure}[htbp]
    \centering
    \includegraphics[height=0.8\linewidth,width=0.8\linewidth]{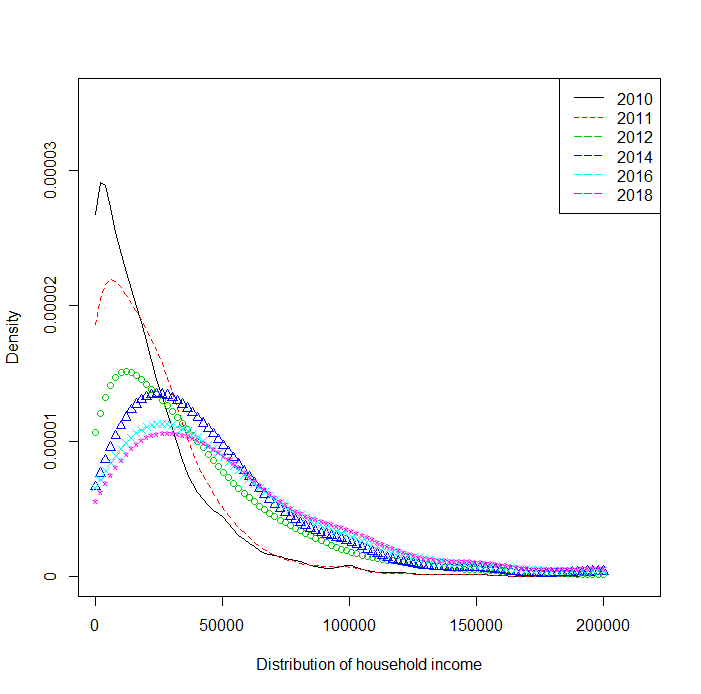}
    \caption{Fitting plot of household income distribution.}
    \label{fig2}
\end{figure}
\subsection{Modeling process}
This paper introduces the distribution functions of education level, family size, income 
structure ratio, and government 
fiscal expenditure as independent variables to model the household income distribution. 
The education level, family size and income structure ratio are independent variables 
with low sampling frequencies (annual), 
while government fiscal expenditure is an independent variable with a high sampling 
frequency (monthly). Therefore, when incorporating government fiscal expenditure as 
an independent variable, the determination of its lag order becomes necessary. In this 
article, the order determination process is as follows. First, the $p$-order lag 
terms of government fiscal expenditure are introduced,
\begin{equation}
    f_{t}(x)=\sum_{k=1}^{3}a_{k}h_{k,t}(x)+a_{4}\sum_{i=1}^{p}b(i,\Theta)g_{t-i/m}(x)+e_{t}(x),
\end{equation}
where, $h_{1,t}(x)$, $h_{2,t}(x)$, $h_{3,t}(x)$, and $g_{t}(x)$ represent the distribution functions 
of education level, family size, income structure, and government fiscal expenditure 
respectively. $b(i,\Theta)$ employs the Almon polynomial controlled by two parameters, 
i.e., $\Theta=(\theta_{1},\theta_{2})$. Then, the lag order $p$ is determined using the 
AIC criterion, 
\begin{equation}
    {\rm{AIC}}=2K+T{\rm{ln}}\left(\sum_{t=1}^{T}\sum_{i=1}^{N}
    \left(f_{t}(s_{i})-{\hat{f}}_{t}(s_{i})\right)^2\Delta s/T\right),
\end{equation} 
where, $K$ is the number of unknown parameters. As shown in Figure 3, the AIC value 
reaches the minimum when the lag order $p=12$. Hence, we introduce government 
fiscal expenditure lagged by $12$ orders as the independent variable. 
\begin{figure}[htbp]
    \centering
    \includegraphics[height=0.6\linewidth,width=0.6\linewidth]{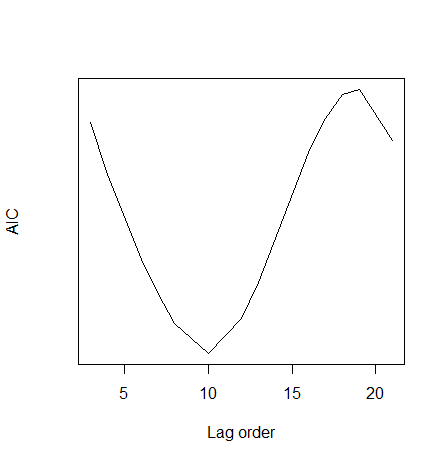}
    \caption{AIC values for different lag orders.}
    \label{fig3}
\end{figure}
\subsection{Model testing}
After incorporating the independent variables, it is crucial to assess the significance 
of the impact of education level, family size, income structure, and government fiscal 
expenditure. Therefore, a significance test needs to be conducted on the regression 
coefficients $a_{1}$ and $a_{2}$ of the PDF-MIDAS model.

The conventional significance testing method may not be suitable for the model employed 
in this article. Therefore, the non-parametric Bootstrap method is utilized to test 
whether the regression coefficient $a_{i}$ is equal to zero. The null hypothesis is 
defined as $a_{i} = 0$, while the alternative hypothesis is $a_{i} \neq 0$. The 
following Bootstrap procedure can estimate the p-value.

\noindent
{\textbf{Step 1:}} Define ${\hat{\varepsilon}}_{t}(s_{i})=f_{t}(s_{i})-{\hat{f}}_{t}(s_{i})$, $t=1,\ldots,T$, $i=1,\ldots,N$.

\noindent
{\textbf{Step 2:}} Generate Bootstrap sample $f_{t}^{(b)}(s_{i})$, 
\[f_{t}^{(b)}(s_{i})={\hat{f}}_{t}(s_{i})+{\hat{\varepsilon}}_{t}^{(b)}(s_{i})\], 
where, $b=1,\ldots,B$. $B$ represents the total number of Bootstrap samples. 
${\hat{\varepsilon}}_{t}^{(b)}(s_{i})$ is independently drawn with replacement from 
$\{{\hat{\varepsilon}}_{t}^{(b)}(s_{1}), \ldots, {\hat{\varepsilon}}_{t}^{(b)}(s_{N})\}$. 

\noindent
{\textbf{Step 3:}} For each Bootstrap sample $b$, the regression coefficient 
${\hat{a}}_{i}^{(b)}$ can be estimated. 

By employing the Bootstrap process, we can obtain the empirical distribution function 
of ${\hat{a}}_{i}$ under the assumption that the null hypothesis is true. Subsequently, 
we can calculate the p-value based on this empirical distribution function.
\begin{equation}
    p-value=\frac{1}{B}\sum_{b=1}^{B}I_{[{\hat{a}}_{i},\infty]}({\hat{a}}_{i}^{(b)}).
\end{equation}
Table 8 presents the estimated values and p-values of the regression coefficients $a_{k}$ 
in model (11). The impact of the PDF of education level, income structure, and local 
government fiscal expenditure on the PDF of household income is significant at a 
significance level of $0.05$.
\begin{table}
    \centering
    \caption{Estimated coefficients and test P values of PDF-MIDAS model. }
    \begin{tabular}{ccc}
    \toprule
    Coefficient  & Estimated value & p-value\\
    \midrule
     $a_{1}$ & 0.143 &0.001\\
     $a_{2}$ & 0.001 &0.997\\
     $a_{3}$ & 0.101 &0.002\\
     $a_{4}$ & 0.755 &0.001\\
    \bottomrule
    \end{tabular}
\end{table}
\subsection{Prediction}
To validate the effectiveness of PDF-MIDAS, we selected PDF-UMIDAS and AVE as 
benchmark models for comparison in our experiments with actual household income data. 
An introduction to each model used as a baseline is shown below. 

(1) The PDF prediction model based on U-MIDAS (Foroni et al., 2015) is abbreviated as 
PDF-UMIDAS. This model does not constrain the weight function form of high-frequency 
independent variables, and only needs to satisfy $\sum_{k=1}^{3}a_{i}+\sum_{i=1}^{p}c_{i}=1$,
\[
   f_{t}(x)=\sum_{k=1}^{3}a_{k}h_{k,t}(x)+\sum_{i=1}^{p}c_{i}g_{t-i/m}(x)+e_{t}(x),  
\]
where, the empirical results of PDF-UMIDAS reach the optimal when the lag order $p=8$. 

(2) The average estimation method (AVE) expresses the forecast value of the current 
period as the average of all lagged periods, 
\[
    f_{t}(x)=\frac{1}{t}\sum_{i=1}^{t}f_{t-i}(x)+e_{t}(x).
\] 

This article conducts a comparative analysis of the prediction effects of various 
models on household income distribution in 2020. The models are trained using actual 
data from 2010 to 2018 as a training set. The evaluation of the prediction 
performance is based on the out-of-sample prediction error, which is measured using 
the mean squared error (MSE), 
\[
    MSE=\|f_{2020}(x)-{\hat{f}}_{2020}(x)\|_{2}^{2}.
\]
In addition to MSE, this paper also measures the structural difference between the 
predicted distribution and the true distribution using the Wasserstein distance. The 
results are presented in Table 9. Figure 4 illustrates the predictions of the three 
models for the household income distribution in 2020, with the black curve representing 
the true distribution function. Furthermore, Table 10 compares the moment information 
between the predicted distribution obtained from the three 
prediction models and the true distribution. 
\begin{table}
    \centering
    \caption{Out-of-sample prediction errors of each model for household income distribution in 2020. }
    \begin{adjustbox}{width=0.7\textwidth}
    \begin{tabular}{cccc}
    \toprule
      & PDF-MIDAS & PDF-UMIDAS &AVE\\
    \midrule
     MSE & {\textbf{0.0038}} &0.0167 &0.0576\\
     Wasserstein Distance & {\textbf{0.0476}} &0.0663 &0.1809\\
    \bottomrule
    \end{tabular}
\end{adjustbox}
\end{table}
\begin{table}
    \centering
    \caption{Moment information of predicted distribution of household income in 2020.  
    TRUE indicates the real household income distribution in 2020. The numbers in 
    parentheses represent the absolute difference from the statistics calculated on 
    the true income distribution.}
    \begin{adjustbox}{width=1\textwidth}
    \begin{tabular}{cccccccc}
    \toprule
      & Mean & SD &$25\%$ &Median & $75\%$ &Skewness & kurtosis \\
    \midrule
     TRUE & 80282 &80111 &33446&63520 &115907 &2.28 &6.76\\
     PDF-MIDAS & {\textbf{81063}} &{\textbf{71883}} &{\textbf{22611}}& {\textbf{56260}} &{\textbf{105172}} &{\textbf{2.09}} &{\textbf{4.79}}\\
     & {\textbf{(781)}} &{\textbf{(8228)}} &{\textbf{(10835)}} &{\textbf{(7260)}} &{\textbf{(10735)}} & {\textbf{(0.19)}} &{\textbf{(1.97)}}\\
     PDF-UMIDAS & 144356 &118906 &48879 &84532 &136647 &0.91 &0.01\\
     &(64074)& (38795) & (15433) &(21012) &(20740) &(1.37) &(6.75)\\
     AVE&46652 &47415 &17213 &37037 &75473 &3.20 &16.46\\
     &(33630)& (32696) & (16233) &(26483) &(40434) &(0.92) &(9.70)\\
    \bottomrule
    \end{tabular}
\end{adjustbox}
\end{table}
\begin{figure}[htbp]
    \centering
    \includegraphics[height=0.8\linewidth,width=0.8\linewidth]{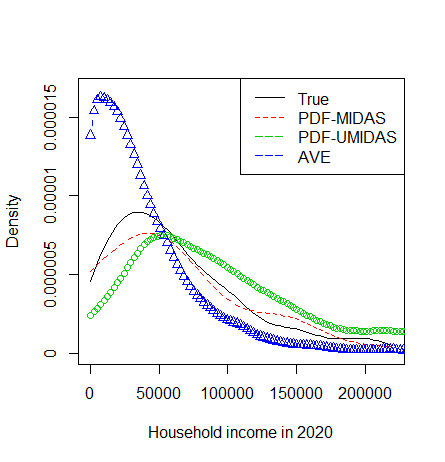}
    \caption{Predictions of household income distribution by various models in 2020.}
    \label{fig4}
\end{figure}

Table 9 and Figure 4 clearly demonstrate that the PDF-MIDAS method proposed in this 
article exhibits the best prediction performance for household income distribution 
in 2020. In terms of both MSE and Wasserstein distance, the PDF-MIDAS method 
showcases a significant improvement compared to the other two methods. 
Figure 4 illustrates that the household income distribution predicted by the PDF-MIDAS 
method in 2020 closely resembles the actual income distribution. Both distributions 
exhibit a unimodal shape, and the position and probability density level of the peak 
in the predicted distribution are the closest to those of the actual distribution. 
On the other hand, the predicted distributions obtained by the PDF-UMIDAS and AVE 
methods show significant discrepancies in terms of both the position of the single 
peak and the probability density level. Furthermore, the true income distribution is 
characterized as a right-skewed distribution. The right-skewed distribution predicted 
by the PDF-MIDAS method closely approximates the true distribution, whereas other 
prediction models exhibit larger deviations. Table 10 provides evidence that the 
moment information of the household income distribution in 2020 predicted by the 
PDF-MIDAS method is the closest to the actual distribution, with the smallest absolute 
difference. For instance, the skewness of the household income distribution in 2020 
is 2.28, while the predicted distribution obtained by the PDF-MIDAS method has a 
skewness of 2.13, indicating a close match. Conversely, the predicted distributions 
obtained by the PDF-UMIDAS and AVE methods demonstrate larger differences in skewness. 

To further illustrate the prediction effect of PDF-MIDAS, this article extends the 
predictions to household income distributions in 2013, 2015, 2017, and 2019. As CFPS 
has not released real household income distribution data for these years, alternative 
indicators are utilized. We utilize the per capita disposable income of residents, 
median per capita disposable income of residents, per capita disposable income of 
low-income, lower middle-income, middle-income, upper middle-income, and high-income 
households, as released by the National Bureau of Statistics of China. We express 
these indicators as household income by multiplying them by $2$. 
Table 11 displays the prediction effects of the PDF-MIDAS, PDF-UMIDAS, and AVE methods 
on each indicator. Even for the four years with missing data, the 
PDF-MIDAS method demonstrates superior prediction performance, with minimal 
discrepancies from the true income distribution characteristics. For instance, in the 
case of average household income in 2019, the absolute error of PDF-MIDAS is only 
1434, significantly smaller than the 43674 of PDF-UMIDAS and the 17123 of AVE. 
Additionally, PDF-MIDAS exhibits the smallest absolute error for the median of the 
household income distribution. These findings indicate that the PDF-MIDAS method can 
accurately capture the evolving characteristics of household income distribution over 
time. Consequently, it plays a crucial role in facilitating further in-depth research 
on the income distribution of Chinese households.
\begin{table}
    \centering
    \caption{Characteristics of the predicted distribution of household income from 
    2013 to 2019. TRUE indicates various characteristics of real household income distribution. 
    The numbers in parentheses represent the absolute difference from the statistics calculated on 
    the true income distribution.}
    \begin{adjustbox}{width=1\textwidth}
    \begin{tabular}{ccccccccc}
    \toprule
     Year& Model& Mean & Low &Lower middle &Middle & Upper middle &High & Median \\
    \midrule
     2013&TRUE & 36622 &8804 &19308&31396 &48722 &94914 &31264\\
     &PDF-MIDAS & {\textbf{35574}} &{\textbf{6741}} &18074&{\textbf{32049}} & {\textbf{50251}} &87655 &{\textbf{27286}}\\
     && {\textbf{(1048)}} &{\textbf{(2063)}} &(1234) &{\textbf{(653)}} &{\textbf{(1529)}} & (7259) &{\textbf{(3978)}}\\
     &PDF-UMIDAS & 46743 &6704 &{\textbf{18458}} &32961 &52030 &88307 &41447\\
     &&(10121)& (2100) & {\textbf{(850)}} &(1565) &(3308) &(6607) &(10183)\\
     &AVE&29184 &6477 &17995 &32305 &51165 &{\textbf{88321}} &27158\\
     &&(7438)& (2327) & (1313) &(909) &(2443) &{\textbf{(6593)}} &(4106)\\
     \midrule
     2015&TRUE & 43932 &10442 &23788&38640 &58876 &109088 &38562\\
     &PDF-MIDAS & 36368 &9045 &21716&{\textbf{38498}} & 62270 &116443 &{\textbf{34034}}\\
     && (7564) &(1397) &(2072) &{\textbf{(142)}} &(3394) & (7355) &{\textbf{(4528)}}\\
     &PDF-UMIDAS & 69393 &{\textbf{9514}} &{\textbf{22823}} &40173 &63380 &120494 &62791\\
     &&(25461)& {\textbf{(928)}} & {\textbf{(965)}} &(1533) &(4504) &(11406) &(24229)\\
     &AVE&{\textbf{37450}} &9237 &22118 &38963 &{\textbf{61703}} &{\textbf{113466}} &30816\\
     &&{\textbf{(6482)}}& (1205) & (1670) &(323) &{\textbf{(2827)}} &{\textbf{(4378)}} &(7746)\\
     \midrule
     2017&TRUE & 51948 &11916 &27686&44990 &69074 &129868 &44816\\
     &PDF-MIDAS & {\textbf{52241}} &10794 &23502&39540 & 66144 &{\textbf{128103}} &{\textbf{49170}}\\
     && {\textbf{(293)}} &(1122) &(4184) &(5450) &(2930) & {\textbf{(1765)}} &{\textbf{(4354)}}\\
     &PDF-UMIDAS & 74836 &{\textbf{11043}} &{\textbf{23821}} &{\textbf{40107}} &{\textbf{67575}} &136151 &69596\\
     &&(22888)& {\textbf{(873)}} & {\textbf{(3865)}} &{\textbf{(4883)}} &{\textbf{(1499)}} &(6283) &(24780)\\
     &AVE&42001 &10449 &23248 &39103 &65627 &124921 &34178\\
     &&(9947)& (1467) & (4438) &(5887) &(3447) &(4947) &(10638)\\
     \midrule
     2019&TRUE & 61466 &14760 &31554&50070 &78460 &152802 &55080\\
     &PDF-MIDAS & {\textbf{60032}} &12990 &30098&49342 & {\textbf{77530}} &{\textbf{149700}} &{\textbf{59385}}\\
     && {\textbf{(1434)}} &(1770) &(1456) &(728) &{\textbf{(930)}} & {\textbf{(3102)}} &{\textbf{(4305)}}\\
     &PDF-UMIDAS & 105140 &{\textbf{13279}} &{\textbf{30530}} &{\textbf{50180}} &79435 &162322 &101757\\
     &&(43674)& {\textbf{(1481)}} & {\textbf{(1024)}} &{\textbf{(110)}} &(975) &(9520) &(46677)\\
     &AVE&44343 &12443 &29332 &49147 &76663 &142736 &37037\\
     &&(17123)& (2317) & (2222) &(923) &(1797) &(10066) &(18043)\\
    \bottomrule
    \end{tabular}
\end{adjustbox}
\end{table}

\section{Conclusion}
In modern time series analysis, dealing with a large number of numerical observations 
at each time point has become a significant research topic. One approach to handling 
such data is to transform the numerous observations into a probability density 
function and conduct statistical modeling. Notable studies in this area include 
Tsay's (2016) HDAR model and Chaudhuri et al.'s (2016) FAR model. 
However, in the context of income distribution prediction, these two models (HDAR and 
FAR) do not take into account high-frequency observed exogenous variables. Additionally, 
the FAR model cannot be directly applied due to the discontinuous sampling time of 
income distribution data. In response to this limitation, this paper introduces a 
novel approach that combines the HDAR and MIDAS models, resulting in the development 
of a mixed data sampling regression model for probability density functions (PDF-MIDAS).  
Given that high-frequency observed independent variables and their lag terms can 
introduce a large number of parameters, simplifying the parameter structure becomes a 
crucial aspect of modeling. This work addresses this concern by employing exponential 
Almon polynomials, which have fewer parameters. These polynomials help control the 
coefficients associated with the lag terms of the predicted independent variables, 
effectively mitigating the issue of parameter expansion that arises in high-dimensional 
scenarios. This article also provides an analysis of the properties of nonlinear 
least squares estimators. Simulation analysis indicates that both univariate and 
multivariate PDF-MIDAS models demonstrate improved performance as the number of 
cross-sectional samples $M$ and the observation length $T$ increase simultaneously. 
Furthermore, when analyzing real data, the PDF-MIDAS model outperforms both the AVE 
and the PDF-UMIDAS models in terms of prediction accuracy. 

Further research is warranted on the following aspects. Firstly, the PDF-MIDAS model 
relies on managing the potential high-dimensionality issue of lag term coefficients. 
It is necessary to explore simplified structures that can reduce the 
complexity of model. In this regard, one possible approach is to express the weight coefficient 
$c_i$ in equation (1) as an expansion using a basis function, 
\[
    c_{i}=\sum_{l=1}^{L}a_{l}w_{l}[(i-1)/m],
\]
where $w_{l}$ is a specific set of basis functions, with the power series being the 
most commonly used option. By utilizing this method, the weight coefficient can be 
controlled through a small number of coefficients $a_l$. The parameter structure can 
then be simplified by applying penalty to the $a_l$ coefficients. Secondly, the 
PDF-MIDAS model proposed in this article can be seen as an extension of the FAR model, 
but with significantly simplified parameters. By considering $f_t(x)$ as an $N\times 1$ 
dimensional vector, we can establish the following model,
\[
   f_{t}(x)=\sum_{i=1}^{s}A_{i}f_{t-s}(x)+\sum_{i=1}^{p}B_{i}g_{t-h-i/m}^{(m)}(x)+e_{t(x)}, 
\]
where, $A_{i}$ and $B_{i}$ are square matrices. To address the challenge of a large 
number of estimated parameters in this model, we can employ reduced rank estimation 
or diagonal matrix estimation for $A_{i}$ and $B_{i}$ to avoid the parameter 
expansion of the model. 
\section*{Acknowledgements}
This research was funded by the Youth Project of 
National Social Science Fund of China (grant 21CTJ008).
Sincere gratitude should go to Digital Economy 
Laboratory in University of International Business 
and Economic for their providing computational 
resources.
\appendix
\renewcommand{\theequation}{A.\arabic{equation}} 
\setcounter{equation}{0} 
\section{Proof of Theorem 1.}
Proof. First we define $G(g_{t,i},\Phi)=\sum_{k=1}^{K}a_{k}B_{k}(L^{1/m_{k}},\Theta_{k})g_{t-h,k}^{(m_{k})}(s_{i})$, 
$g_{t,i}=(g_{t-h-1/{m_{1}},1}^{(m_{1})}(s_{i}),\ldots,g_{t-h-p_{K}/m_{K},K}^{(m_{k})}
(s_{i}))$, and the 2-norm of vector $x$ is $\|x\|_{2}=\sqrt{x^{\top}x}$. 
The $\triangledown$ represents differentiation.  
Without loss of generality, we assume that the number of parameters of the weight 
function $q_{1}=\ldots=q_{K}=1$. 
According to Theorem 4.1 and $\|\cdot\|_{s}$ norm of Mira and Escribano (1995), for any $C>0$, it 
is sufficient to demonstrate that the following conditions are satisfied in order to 
prove Theorem 1. 

\noindent
{\textbf{Condition 1:}} $|G(g_{t,i},\Phi)|\le C\|g_{t,i}\|_{s}$, 

\noindent
{\textbf{Condition 2:}} For the vector norms $\|\cdot\|_{s}$ and $\|\cdot\|_{2}$, 
\[\|{\triangledown}_{g_{t,i}}G(g_{t,i},\Phi)\|\le C,\]

\noindent
{\textbf{Condition 3:}} $\|{\triangledown}_{\Phi}G(g_{t,i},\Phi)\|_{s}^{2}\le C\|g_{t,i}\|_{s}^{2}$, 

\noindent
{\textbf{Condition 4:}} For $j$, $l=1,\ldots,2K$, 
\[\|{\triangledown}_{\Phi}\frac{\partial}{\partial \Phi_{j}}G(g_{t,i},\Phi)\|_{s}^{2}\le C\|g_{t,i}\|_{s}^{2}, \ 
\|{\triangledown}_{\Phi}\frac{\partial^{2}}{\partial \Phi_{j} \partial \Phi_{l}}G(g_{t,i},\Phi)\|_{s}^{2}\le C\|g_{t,i}\|_{s}^{2},\]
where, $\Phi_{j}$ represents the $j$th parameter in $\Phi$. 

For Condition 1, because $a_{k}$ and $c_{i,k}$ in (4), $k=1,\ldots,K$, $i=1,\ldots,p_{k}$ are bounded,  
\[
    |G(g_{t,i},\Phi)|=|\sum_{k=1}^{K}a_{k}B_{k}(L^{1/m_{k}},\Theta_{k})g_{t-h,k}^{(m_{k})}(s_{i})|\le C\|g_{t,i}\|_{s}, 
\]
holds. For Condition 2, we have 
\[
   \triangledown_{g_{t,i}}G(g_{t,i},\Phi)=(a_{1}c_{1,1},\ldots,a_{1}c_{p_{1},1},\ldots,a_{K}c_{1,K},\ldots,a_{K}c_{p_{K},K})^{\top},
\]
so $\|{\triangledown}_{g_{t,i}}G(g_{t,i},\Phi)\|\le C$ is obviously established. 

For Condition 3 and 4, we have 
\[
   \left\{
    \begin{aligned}
      &\frac{\partial}{\partial a_{k}} G(g_{t,i},\Phi)=\sum_{i=1}^{p_{k}}c_{i,k}g_{t-h-i/m_{k}}^{(m_{k})}(x), \\
      &\frac{\partial}{\partial \theta_{k}} G(g_{t,i},\Phi)=a_{k}\sum_{i=1}^{p_{k}}\frac{\partial c_{i,k}}{\partial \theta_{k}}g_{t-h-i/m_{k}}^{(m_{k})}(x), \\
    \end{aligned}
    \right. 
\]
\[
   \left\{
    \begin{aligned}
      &\frac{\partial^2}{\partial \theta_{k}\partial a_{k}} G(g_{t,i},\Phi)=\sum_{i=1}^{p_{k}}\frac{\partial c_{i,k}}{\partial \theta_{k}}g_{t-h-i/m_{k}}^{(m_{k})}(x), \\
      &\frac{\partial^2}{\partial \theta_{k}^{2}} G(g_{t,i},\Phi)=a_{k}\sum_{i=1}^{p_{k}}\frac{\partial^{2} c_{i,k}}{\partial \theta_{k}^{2}}g_{t-h-i/m_{k}}^{(m_{k})}(x), \\
    \end{aligned}
    \right. 
\]
\[
   \left\{
    \begin{aligned}
      &\frac{\partial^3}{\partial \theta_{k}^{2}\partial a_{k}} G(g_{t,i},\Phi)=\sum_{i=1}^{p_{k}}\frac{\partial^{2} c_{i,k}}{\partial \theta_{k}^{2}}g_{t-h-i/m_{k}}^{(m_{k})}(x), \\
      &\frac{\partial^3}{\partial \theta_{k}^{3}} G(g_{t,i},\Phi)=a_{k}\sum_{i=1}^{p_{k}}\frac{\partial^{3} c_{i,k}}{\partial \theta_{k}^{3}}g_{t-h-i/m_{k}}^{(m_{k})}(x). \\
    \end{aligned}
    \right. 
\]
Because $c_{i,k}$, $\frac{\partial c_{i,k}}{\partial \theta_{k}}$, 
$a_{k}\frac{\partial c_{i,k}}{\partial \theta_{k}}$, 
$\frac{\partial^{2} c_{i,k}}{\partial \theta_{k}^{2}}$, 
$a_{k}\frac{\partial^{2} c_{i,k}}{\partial \theta_{k}^{2}}$ and 
$\frac{\partial^{3} c_{i,k}}{\partial \theta_{k}^{3}}$ are bounded, Conditions 3 and 
4 hold. According to Theorem 4.1 of Mira and Escribano (1995), the asymptotic normality 
of the estimator ${\hat{\Phi}}$ can be achieved. 



\end{document}